# STRINGS, TOPOLOGICAL CHANGE
# AND
# DARK MATTER

**Gerald E. Marsh**
**Argonne National Laboratory (Retired)**
**gemarsh@uchicago.edu**

**ABSTRACT**

Dark matter, first postulated by Jacobus Kapteyn in 1922 and later by Fritz Zwicky in 1933, has remained an enigma ever since proof of its existence was confirmed in 1970 by Vera Rubin and Kent Ford by plotting the rotation curve for the Andromeda galaxy. Here, some concepts from string theory and topological change in quantum cosmology are used to formulate a new model for dark matter. The density profiles of dark matter halos are often modeled as an approximate solution to the Lane-Emden equation. Using the model proposed here for dark matter, coupled with previous work showing that the approximate solution to the Lane-Emden equation can be an exact solution of the Einstein-Maxwell equations, provides a new insight into the possible nature of dark matter.

# Introduction

Dark matter, first postulated by Jacobus Kapteyn in 1922 and later by Fritz Zwicky in 1933, has remained an enigma ever since proof of its existence was confirmed in 1970 by Vera Rubin and Kent Ford by plotting the rotation curve for the Andromeda galaxy. Two iconic images related to dark matter are shown below.

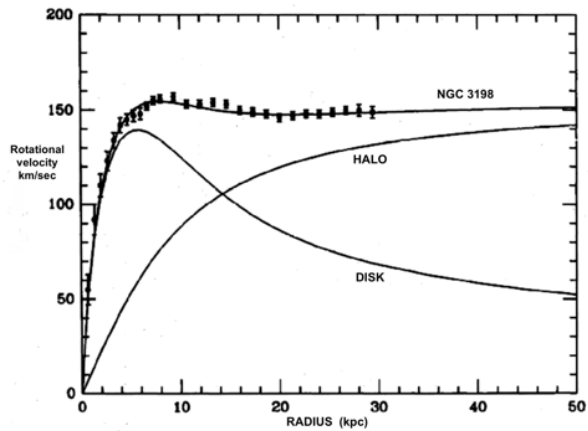 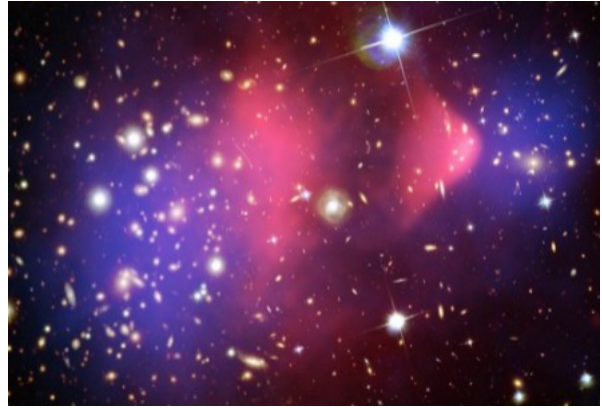

Rotation curve [Adapted from T.S. Albada, et al., "Distribution of Dark Matter in the Spiral Galaxy NGC 3198", ApJ, **295**, 305 (1985)]; Bullet Cluster (1E0657-56) Composite Credit: X-ray: NASA/CXC/CfA/ M.Markevitch et al.; Lensing Map: NASA/STScI; ESO WFI; Magellan/U.Arizona/ D.Clowe et al. Optical: NASA/STScI; Magellan/U.Arizona/D.Clowe et al.

The rotation curve for NGC 3198 shows that the velocity of visible matter is essentially flat for distances greater than ~5 kpc from the center of the galaxy, instead of having a Keplerian fall-off proportional to $1/r$ (See the ApJ paper for a discussion of the spherical halo and exponential disk). The composite image on the right shows the relatively recent collision of two galaxy clusters. The two pink areas contain most of the ordinary mass of the two clusters, the bullet-shaped one having passed through the other larger cluster. In the process of the collision, the temperature of the normal matter is increased and X-rays are emitted that were detected by the Chandra X-Ray Observatory. The blue areas are a map of the invisible matter made by using gravitational lensing, where light from objects more distant than the bullet cluster is bent by intervening matter. The normal matter shown in pink is clearly separate from the majority of the matter comprising the clusters shown in blue. The conclusion being that most of the matter in the clusters is dark matter.

In this paper, some concepts from string theory, along with the possibility of topological change through quantum tunneling, are used to construct a scenario for the evolution of the early universe and possibly give some insight into the nature of dark matter. The scenario envisions the very early universe as a 3-sphere that plays the role of a brane in string theory where the ends of open strings bearing a Kalb-Ramond charge are terminated. As the universe expands, still in its early phase, its topology changes from being a positively curved 3-sphere to being negatively curved, which is consistent with recent data showing that the universe may indeed be negatively curved. While such a topological change would classically imply the appearance of acausal features, that need not be the case for quantum topological transitions in the early universe. The possibility that the charged end points of terminated strings can play the role of dark matter is discussed and it is shown that such dark matter gives an exact solution to the Einstein-Maxwell equations that matches the density profiles of dark matter halos that are generally modeled as an approximate solution to the Lane-Emden equation.

Section 1 introduces some features from string theory; Section 2 discusses D3-branes and Friedmann-Lemaître-Robinson-Walker cosmological models; the Kalb-Ramond charged string terminating in $\mathbb{S}^3$ is treated in Section 3; Section 4 discusses a scenario for the appearance of only one sign of "dark" charges in $\mathbb{S}^3$; the evolution of the universe and topological change is covered in Section 5; Section 6 gives a string model for dark matter; and Section 7 discusses dark matter as charged dust now based on string theory.

**1. Some concepts from string theory**

There are many excellent books on string theory that would expand on this limited conceptual introduction. Two of the more accessible are by Zwiebach[1] and Tong[2]. String theory uses a (D+1)-dimensional Minkowski space with D spatial dimensions.

The Kalb-Ramond massless antisymmetric gauge field $B_{\mu\nu} = -B_{\nu\mu}$, can be viewed as the *analog* of the Maxwell gauge field $A_\mu$ of electromagnetics. In the case of electromagnetism, the field strength is given by $F_{\mu\nu} = \partial_\mu A_\nu - \partial_\nu A_\mu$.

For $B_{\mu\nu}$ the field strength, $H_{\mu\nu\rho}$, is defined as $H_{\mu\nu\rho} = \partial_\mu B_{\nu\rho} + \partial_\nu B_{\rho\mu} + \partial_\rho B_{\mu\nu}$, $H_{\mu\nu\rho}$ being a totally antisymmetric tensor corresponding to a torsion field (for further discussion, see Appendix A). The theory that allows space-time to have torsion is the Einstein-Cartan theory, which is a modification of general relativity that in the cosmological context would have only a very slight effect.[3] The usual Einstein-Hilbert action is

$$S_{E-H} = \frac{1}{16\pi G} \int d^4x \sqrt{-g}\, R$$

(1.1)

where $R$ is the scalar curvature and the other symbols have their usual meaning.

If the space allows torsion, the torsion tensor is

$$T_{\alpha\beta}{}^{\mu} = \Gamma_{\alpha\beta}{}^{\mu} - \Gamma_{\beta\alpha}{}^{\mu}.$$

(1.2)

The Einstein-Cartan action in terms of the Riemannian scalar curvature is

$$S = \frac{1}{16\pi G} \int d^4x \sqrt{-g}\, \tilde{R},$$

(1.3)

where

$$\tilde{R} = R + 2T_{\beta;\alpha}{}^{\alpha\beta} - T_{\alpha\beta}{}^{\beta}T_{\gamma}{}^{\alpha\gamma} + \tfrac{1}{4}T_{\alpha\beta\gamma}T^{\alpha\beta\gamma} + \tfrac{1}{2}T_{\alpha\beta\gamma}T^{\gamma\beta\alpha}.$$

(1.4)

For a completely antisymmetric torsion tensor, the Einstein-Cartan action reduces to

$$S_{E-C} = \frac{1}{16\pi G} \int \sqrt{-g}\, d^4x \left(R - H_{\alpha\beta\gamma}H^{\alpha\beta\gamma}\right),$$

$$H_{\alpha\beta\gamma}H^{\alpha\beta\gamma} = -\tfrac{1}{4}T_{\alpha\beta\gamma}T^{\alpha\beta\gamma} - \tfrac{1}{2}T_{\alpha\beta\gamma}T^{\gamma\beta\alpha}.$$

(1.5)

where the field $H_{\mu\nu\rho}$ is derivable from a tensor potential $B_{\mu\nu}$ so that $H_{\mu\nu\rho} = \partial_\mu B_{\nu\rho} + \partial_\nu B_{\rho\mu} + \partial_\rho B_{\mu\nu}$. This $B_{\mu\nu}$ is identical with the Kalb-Ramond antisymmetric tensor field in string theory. As is seen from Eqs. (1.1) and (1.5), when the torsion vanishes the Einstein-Cartan action reduces to the Einstein-Hilbert action.[4] For a completely antisymmetric torsion tensor, the possible metric connections correspond to geodesics that are the same as those derived from a Levi-Civita connection.[5]

In string theory, the general term "D-brane" refers to an "object" upon which string endpoints lie. The letter D stands for the Dirichlet boundary conditions that the endpoint must satisfy on the brane. A D$p$-brane is an object with $p$ spatial dimensions. The general spacetime dimension is $p + 1$, so 4-dimensional spacetime is considered to be a D3-brane. Branes with D spatial dimensions are also called D-branes. D-branes are not necessarily hypersurfaces or of infinite extent, they can also be finite, closed surfaces. The additional spatial dimensions beyond the dimension of the brane are known as comprising the "bulk". It is interesting that Zaslow, in the context of category theory, simply defines branes as "boundary conditions".[6]

The strings of interest here carry Kalb-Ramond string charge. This charge can be viewed as a "current" flowing along the string since the string charge density is a *vector* which is tangent to the string. For 4-dimensional spacetime, the action for the brane and the string will have a $F^{0k}B_{0k}$ term, where $F^{0k}$ comes from the Maxwell field tensor. Since $F^{0k}$ couples to $B_{0k}$ it must carry a string charge, but $F^{0k} = E_k$, so that the field $E_k$ on the brane carries string charge.

The field strength, $H_{\mu\nu\rho}$, as defined above, is totally antisymmetric and invariant under the gauge transformations

$$\delta B_{\mu\nu} = \partial_\mu \Lambda_\nu - \partial_\nu \Lambda_\mu.$$

(1.6)

Here the arguments of $B_{\mu\nu}$ are the string coordinates $X(\tau,\sigma)$. The "world sheet" of an open string is defined as the trajectory of the string in space-time with space-like coordinates $X^\mu$. On this world sheet there are two linearly independent tangent vectors given by $\partial_\tau X^\mu$ and $\partial_\sigma X^\mu$, where $\tau$ parameterizes time and $\sigma$ parameterizes the distance along the string. For bosonic strings, one uses the classical variable $X^\mu(\tau,\sigma)$ to describe the position of the string.

The part of the action that couples the string to the $B_{\mu\nu}$ field is given by

$$S_B = -\frac{1}{2} \int d\tau d\sigma \, \varepsilon^{\alpha\beta} \partial_\alpha X^\mu \partial_\beta X^\nu B_{\mu\nu}.$$

(1.7)

$\varepsilon^{\alpha\beta}$ is totally antisymmetric so that when this action is varied using Eq. (1.6) the result is

$$\delta S_B = -\int d\tau d\sigma (\partial_\tau \Lambda_\nu \, \partial_\sigma X^\nu - \partial_\sigma \Lambda_\nu \, \partial_\tau X^\nu) = -\int d\tau d\sigma (\partial_\tau (\Lambda_\nu \, \partial_\sigma X^\nu) - \partial_\sigma (\Lambda_\nu \, \partial_\tau X^\nu)).$$

(1.8)

Now if $\Lambda$ is set equal to zero at $\pm\infty$,[†] the $\partial_\tau$ term vanishes. Since the open string terminates on a D-brane, the coordinates $X^\mu$ may be divided into those on the brane labeled $X^m$ and those perpendicular to the brane labeled $X^a$. Then integrating Eq. (1.8) on the brane with respect to $\sigma$ gives

$$\delta S_B = \int d\tau \, (\Lambda_m \, \partial_\tau X^m + \Lambda_a \, \partial_\tau X^a) \Big|_{\sigma=0}^{\sigma=\pi}.$$

(1.9)

Because Dirichlet boundary conditions apply at both end points of the string, the term $\Lambda_a \, \partial_\tau X^a$ vanishes when evaluated at these points. Given these boundary conditions, the string ends remain attached and perpendicular to the brane.

For $S_B$ to be gauge invariant $\delta S_B$ must vanish. To make this happen one adds a term to the action coupling the ends of the string to the dark fields on the brane. That is,

$$S = S_B + \int d\tau \, A_m(X) \partial_\tau X^m \Big|_{\sigma=0}^{\sigma=\pi}.$$

(1.10)

For this to work, one must impose the condition $\delta A_m = -\Lambda_m$. Doing so immediately results in $\delta S = 0$ so that gauge invariance is restored. Now, however, since $\delta F_{mn} = -\delta B_{mn}$ neither field is independently gauge invariant. This means that the physical field strength must be redefined as $\mathcal{F}_{mn} = F_{mn} + B_{mn}$. Then on the brane the gauge invariant generalization of the dark field Lagrangian density is $-\frac{1}{4}\mathcal{F}^{mn}\mathcal{F}_{mn}$. Expanding this gives

$$-\frac{1}{4}\mathcal{F}^{mn}\mathcal{F}_{mn} = -\frac{1}{4}B^{mn}B_{mn} - \frac{1}{4}F^{mn}F_{mn} - \frac{1}{2}F^{mn}B_{mn}.$$

(1.11)

---

[†] This type of requirement generally arises from Noether's theorem, which states that every continuous symmetry of the Lagrangian gives rise to a conserved current $j^\mu(x)$ which, when coupled with the equations of motion, implies that $\partial_\mu j^\mu(x) = 0$. The conserved current means that there is a conserved charge $Q = \int_{\mathbb{R}^3} d^3x \, j^0$, provided $\vec{j} \to 0$ quickly enough as $|\vec{x}| \to \infty$.

The last term can be written as

$$-\frac{1}{2}F^{mn}B_{mn} = -F^{0k}B_{0k} + \ldots$$

(1.12)

Since $F^{0k}$ couples to $B_{0k}$ it must carry a string charge, but $F^{0k} = E_k$ so, as mentioned earlier, the dark field on the brane carries string charge.

The real question is how to interpret the second term of Eq. (1.10),

$$\int d\tau\, A_m(X)\partial_\tau X^m \big|_{\sigma=\pi} - \int d\tau\, A_m(X)\partial_\tau X^m \big|_{\sigma=0}.$$

(1.13)

It is generally maintained that these terms add plus and minus charge to the ends of the string; that is, as pointed out in Zwiebach's book, the ends of an open string "behave" as electric point charges. But the first term on the right-hand side of Eq. (1.12) can be interpreted as saying that not only does the electric field on the brane carry string charge, but the string in the "bulk" carries the electric field as well. This is what the $F^{0k}B_{0k}$ term in Eq. (1.12) means—the two fields are coupled: $F^{0k}$ couples to $B_{0k}$ and vice versa. On the brane, the emergence of the electric field looks like a charge; in the bulk the electric field is confined to the string. This is reminiscent of Wheeler's "charge without charge". For a discussion of this concept in the context of string theory, see Marsh.[7]

While it is generally assumed that the charges on the endpoints of strings that terminate on a brane are Maxwell electric charges, this interpretation in terms of electromagnetism is not mandated. Kalb and Ramond did not assume that the charges and fields associated with string endpoints are actually electromagnetic in nature, only that there is an "electromagnetic-*type* interaction between point charges located at the ends of strings" [emphasis added]. The electromagnetic interpretation will not be used here. In what follows, the "apparent" charges of the previous paragraph will be called "dark charges" and their associated fields "dark fields". These fields will nevertheless be assumed to obey the Maxwell equations.

In what follows it is important that the string ends have mass as well as carry dark charges. In string theory, the interaction between strings is generally ignored because in quantum field theory causality requires that point like particles only interact when they overlap in space and time; i.e., there are no interactions at a distance such as the Coulomb interaction in electromagnetics. But here the context is classical and the work of Kalb and Ramond becomes relevant.[8] They use the gauge conditions

$$\partial_\mu B^{\mu\nu} = \frac{g}{e} A^\nu \text{ where } \partial_\nu A^\nu = 0$$

(1.14)

where $g$ is a coupling constant having the dimensions of mass and $e$ is a dimensionless coupling constant, along with the Lagrangian

$$\mathcal{L} = \tfrac{1}{12}H_{\mu\nu\rho}H^{\mu\nu\rho} - \tfrac{1}{4}(\partial_\mu A_\nu - \partial_\nu A_\mu)^2 - \tfrac{1}{4}\left(\tfrac{g}{e}\right)^2 B^{\mu\nu}B_{\mu\nu} - \tfrac{1}{2}\tfrac{g}{e} B^{\mu\nu}(\partial_\mu A_\nu - \partial_\nu A_\mu)$$

(1.15)

to find the equations of motion

$$g^{\mu\nu}\partial_\nu\partial_\mu A^\mu + \left(\tfrac{g}{e}\right)^2 A^\mu = 0.$$

(1.16)

This is the classical Klein-Gordon equation for a massive vector field with mass $g/e$. If one now defines $\mathfrak{B}_{\mu\nu} = B_{\mu\nu} + \tfrac{e}{g}(\partial_\mu A_\nu - \partial_\nu A_\mu)$, the Lagrangian can be written as

$$\mathcal{L} = \tfrac{1}{12}H_{\mu\nu\rho}H^{\mu\nu\rho} - \tfrac{1}{4}\left(\tfrac{g}{e}\right)^2 \mathfrak{B}^{\mu\nu}\mathfrak{B}_{\mu\nu} \text{ where } \partial_\mu \mathfrak{B}^{\mu\nu} = 0.$$

(1.17)

$\mathfrak{B}_{\mu\nu}$ is then a massive pseudovector field. Kalb and Ramond concisely summarize the above as follows: The fields $A_\mu$ and $B_{\mu\nu}$, when taken individually, are associated with massless particles which respectively mediate the long-range forces between open and closed strings. The combined effect of these fields produces a massive pseudovector interaction between open strings.

## 2. D3-branes and Friedmann-Lemaître-Robinson-Walker cosmological models

It has been shown by Lachièze-Rey[9] that all FLRW pseudo-Riemannian manifolds can be embedded in a flat 5-dimensional Minkowski manifold with Lorentzian signature. Every FLRW model is a 4-dimensional submanifold (hypersurface) in this 5-dimensional space.

In what follows, the 3-dimensional spacelike hypersurface or brane will initially be chosen to be the manifold $\mathbb{S}^3$, an oriented manifold that admits a spin structure enabling the existence of spinors. The bulk is a flat space that allows a torsion field.

String theory has branes embedded in a bulk of higher dimension. This bulk must be a torsion space since strings with Kalb-Ramond charge have a completely antisymmetric field intensity that is a torsion field. Thus, the Kalb-Ramond field is a source of torsion. Given the choice of $\mathbb{S}^3$ with a spin structure for the brane, the question is then whether or not a Riemannian manifold can be embedded in a torsion space. The question has been answered affirmatively by Romero, et al.[10]

Trautman[11] has written a paper on the Einstein-Cartan theory that is relevant to the completely antisymmetric nature of the field intensity of the Kalb-Ramond charge. Note that there is no difference between the Einstein-Cartan theory and the Einstein theory when torsion vanishes. In general, Einstein-Cartan theory differs only slightly from General Relativity. The effects only become significant when the spin density squared is comparable to the mass density. Non-zero spin density can only exist in the presence of a medium. Trautman notes that one does not need to introduce torsion to describe spinning matter. Torsion theories can be reformulated as Riemannian theories with an additional torsion tensor that appears as a supplementary term of the energy-momentum tensor in the Einstein field equations.

### *Spin Structures*

Because of the real-world existence of spinors, the D3-brane of the FLRW universe of interest must admit a spin structure. Whether or not this is possible depends on the first two Stiefel-Whitney classes. Following Milnor and Stasheff,[12] some relevant background follows. A concise discussion of spinor bundles has been given by Marsh.[13]

If $S_i^p$, $i = 1, 2, \ldots$ are $p$-simplices they can be taken as the free generators of an abelian group, and a $p$-chain can be defined as

$$C_p = \sum_i g_i S_i^p, \quad g_i \in G,$$

(2.1)

where G is an abelian group. Such $p$-chains also form a group $C_p(K,G)$, $K$ being a topological space. It will be assumed here that $K$ is a manifold. The boundary operator applied to $C_p(K,G)$ yields $\partial : C_p(K,G) \rightarrow C_{p-1}(K,G)$. The kernel of $\partial$ is $Z_p(K,G)$ whose elements are $p$-cycles. The group of bounding $p$-cycles of $K$ over $G$ (boundaries) is then $\partial : C_{p+1}(K,G) \rightarrow B_p(K,G)$. The homology classes are then defined as $H_p(K,G) = Z_p(K,G) / B_p(K,G)$. Similarly, using the coboundary operator $\partial^*$ one can define the cohomology classes as $H^p(K,G) = Z^p(K,G) / B^p(K,G)$. Homology uses the global boundary operator $\partial$, and cohomology uses $\partial^*$, or equivalently the exterior derivative $d$, which is a local operator.

$H^i(K,G)$ is then the $i^{th}$ cohomology group of $K$ with coefficients in $G$. For each vector bundle $\xi$ there corresponds a sequence of cohomology classes

$$w_i(\xi) \in H^i(K(\xi), Z/2), \quad i = 0, 1, 2, \ldots$$

(2.2)

called the Stiefel-Whitney classes of $\xi$. Note that

$$w_0(\xi) = 1 \in H^0(K(\xi), Z/2),$$

(2.3)

and $w_i(\xi) = 0$ for $i > n$ if $\xi$ is an $n$-plane bundle.

The Whitney product theorem requires a definition of the cup product: Given $[\omega] \in H^p(K, \mathbb{R})$ and $[v] \in H^q(K, \mathbb{R})$, $[\omega] \cup [v] = [\omega \wedge v]$ a $p+q$ form, which implies that $[\omega \wedge v] \in H^{p+q}(K, \mathbb{R})$.

Thus, $\cup: H^p(K, \mathbb{R}) \times H^q(K, \mathbb{R}) \to H^{p+q}(K, \mathbb{R})$. If $\xi$ and $\eta$ are vector bundles over the same base space, the Whitney product theorem is

$$w_k(\xi \oplus \eta) = \sum_{i=0}^{k} w_i(\xi) \cup w_{k=i}(\eta).$$

(2.4)

For example,

$$w_1(\xi \oplus \eta) = w_1(\eta) + w_1(\xi)$$
$$w_2(\xi \oplus \eta) = w_2(\eta) + w_2(\xi) + w_1(\xi) \cup w_1(\eta).$$

(2.5)

When a manifold admits spinors it has a spin structure and is known as a spin manifold. Denote the principal bundle of a closed, compact manifold $K$ by $P_{SO(n)}(K)$. $K$ has the structure group SO($n$) and sections of the tangent bundle are vector fields on $K$. If $K$ is a spin manifold, it has a "lifting" of its structure group SO($n$) to the group Spin($n$). A lifting is a principal bundle map from the spin bundle to the frame bundle. Choosing such a lifting constitutes a choice of spin structure on $K$; i.e.,

$$P_{SO(n)}(K) \xrightarrow{lifting} P_{Spin(n)}(K).$$

(2.6)

This also induces a bundle of spinors $S(K)$ over $K$; sections of $S(K)$ are spinors.

While SO($n$) is a connected group it is not simply connected. For $n \geq 3$, the first homotopy group, $\pi_1(SO(n))$, is isomorphic to $\mathbb{Z}_2$. The universal double cover of SO($n$) is the spin group Spin($n$).

The lifting of the structure group SO($n$) to the group Spin($n$) requires that the first two Stiefel-Whitney classes of the tangent bundle of $K$ vanish: $w_1(K) = w_2(K) = 0$. The vanishing of $w_1(K)$ implies that $K$ is orientable and clockwise and anti-clockwise rotations can be distinguished, while the vanishing of $w_2(K)$ makes the double covering of SO($n$) by Spin($n$) global.

It should be noted that the existence of a spin structure on a manifold does not directly relate to the problem of including fermions in string theory. That is addressed by superstring theory, which introduces the idea of supersymmetry between bosons and fermions. This, and subsequent developments in string theory, will play no role in what follows.

*The three-sphere $\mathbb{S}^3$*

$\mathbb{S}^3$ is unique because of the 1904 Poincaré conjecture, which states that a compact, connected 3-manifold is topologically equivalent (homeomorphic) to $\mathbb{S}^3$ if it is simply connected. The conjecture was proved by Grigori Perelman in 2006. Because $\mathbb{S}^3$ will play an important role in what follows, a brief introduction to the different ways of representing $\mathbb{S}^3$ is given in this section.

In general, $\mathbb{S}^n$ is the one-point compactification of $\mathbb{R}^n$; i.e., $\mathbb{S}^n = \mathbb{R}^n \cup \{\infty\}$. The model of $\mathbb{S}^3$ that will be used here is the identification of the boundary of two balls $B^3{}_1, B^3{}_2$ by a homeomorphism $h$ of $\partial B^3{}_1$ onto $\partial B^3{}_2$. This is shown in Fig. 2.1.

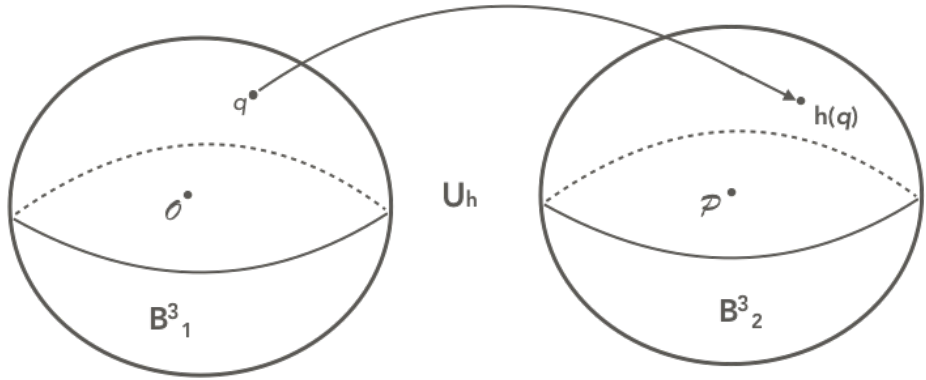

Figure 2.1. The two-ball model of $\mathbb{S}^3$ given by the union of the surface of two 3-balls given by $h: \partial B^3{}_1 \to \partial B^3{}_2$ so that $\mathbb{S}^3 = B^3{}_1 \cup_h B^3{}_2$. The point $q$ is on the surface (boundary) of $B^3{}_1$ and $h(q)$ is on the surface of $B^3{}_2$.

If $\mathcal{P} \subset B^3{}_2$ is set equal to $\infty$, as will later be the case, then $B^3{}_2 = \{\mathcal{P}_\infty\} \cup (\mathbb{R}^3 - \text{Int } B^3{}_1)$, where "Int" means interior. $\{\mathcal{P}_\infty\} \cup (\mathbb{R}^3 - \text{Int } B^3{}_1)$ is a topological ball with center at infinity. This is shown in Figs. 2.2(a) and 2.2(b) for a 2-dimensional projection. The 1-point compactification of $\mathbb{R}^3$ (i.e., $\mathbb{S}^3$) is then the union of this $B^3{}_2$ with $\mathcal{P}$ at infinity and $B^3{}_1$ sewn together by the identity map of $\partial B^3{}_1$ onto $\partial B^3{}_2$. This model of $\mathbb{S}^3$ is equivalent to that shown in Fig. 2.1.[14, 15]

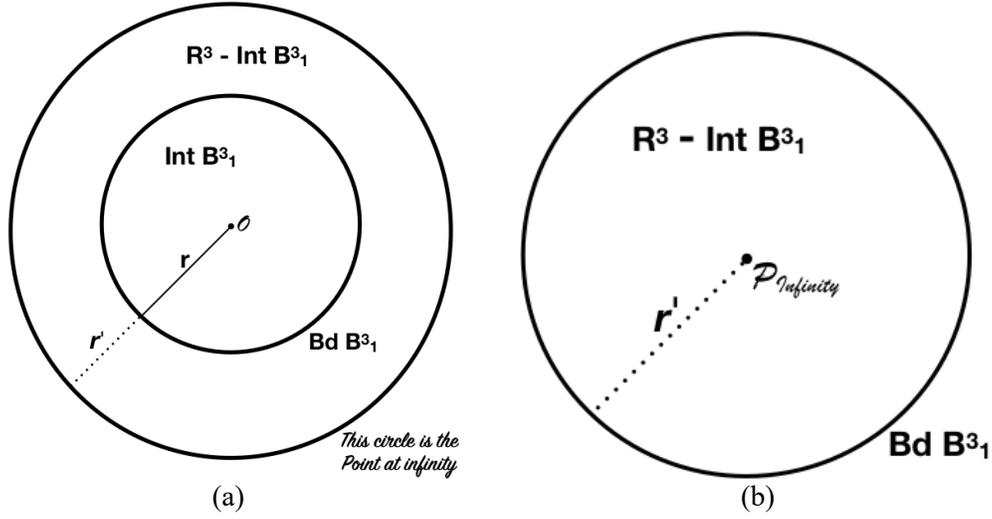

Figure 2.2. (a) A 2-dimensional projection of the 3-ball $B^3_1$; (b) 3-Ball $B^3_2$ when P $\to \infty$. The radius r′ of $B^3_2$ is given by $\{P_\infty\} \cup$ r′.

## 3. Kalb-Ramond charged string terminating in $\mathbb{S}^3$

The presence of dark charges within $\mathbb{S}^3$ raises the question of whether or not a single-valued $B$ field over $\mathbb{S}^3$ can exist. Here, the dark charges at the end of open strings will be isolated by additional boundary components. These affect the topology of the space. The isolation of the charges associated with a Kalb-Ramond field by boundary components has also been used by Bowick, et al.[16]

Strings that end on branes are generally discussed from the perspective of the bulk; often, for example, by a string terminating on a D2-brane. The charged strings here terminate on $\mathbb{S}^3$ and from the perspective of the interior of $\mathbb{S}^3$ the end charges appear as individual points that are isolated from the surrounding space by internal boundary components as shown in Fig. 3.1. The string itself is contained in the bulk within which $\mathbb{S}^3$ is situated, while the field from the dark charges is contained in $\mathbb{S}^3$ and does not enter the bulk (Zwiebach, §15.3) itself, but—as discussed in section 1—are constrained to the string.

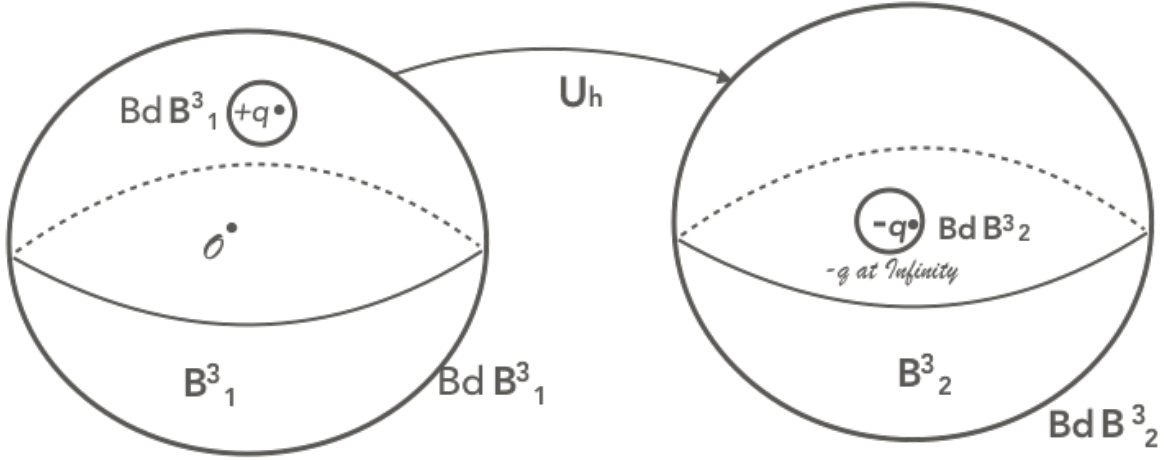

Figure 3.1. The two-ball model of $\mathbb{S}^3$ given by $\mathbb{S}^3 = B^3_1 \cup_h B^3_2$ with a Kalb-Ramond charged string with the dark charges at each end of opposite sign. The dark charges are *within* their respective 3-balls. These are surrounded by interior boundary components $Bd\ B^3_1$ and $Bd\ B^3_2$ that isolate the charges. The radius of these interior boundary components is that of the charge; e.g., zero for a point charge. $P \subset B^3_2$ has been set at $\infty$. The topology near the charges located at $p_i$ is then locally equivalent to $\mathbb{R}^3 - \{p_i\}$.

The Kalb-Ramond antisymmetric gauge field, or potential, $B_{\mu\nu}$ looks like an electromagnetic potential with an additional index, and transforms under a gauge transformation in the same way. $B_{\mu\nu}$ can be viewed as a 2-form $B$, so that $H = dB$. The 3-form $H$ is closed since $dH = 0$. The 3-form $H$ will be exact (so that the potential $B_{\mu\nu}$ is globally defined) if and only if the third homology group $H_3(\mathbb{S}^3)$ vanishes; i.e., if and only if the period $\int_{\partial \mathbb{M}} H = 0$. This is not the case for $\mathbb{S}^3$ but *is* the case once there are boundary components isolating the dark charges at each end of the string. In summary, for a single-valued potential $B$ to exist, the third homology group of $\mathbb{S}^3$ must vanish. Appendix B elaborates on these issues using the example of $\mathbb{S}^2$.

## 4. Only one sign for dark charges in $\mathbb{S}^3$.

In what follows, it is necessary that only one sign of the dark charge at the ends of Kalb-Ramond strings appear within $\mathbb{S}^3$. That there exists at least one possible scenario for this to occur will be shown here. It is based on the idea of cosmic strings and inflation in the early history of the universe.

As the Universe cooled since the Planck time there have been a series of spontaneous symmetry breaking phase transitions during which topological defects, such as cosmic strings,[17] could have formed. An example of a string-like topological defect is the magnetic flux line in a type II superconductor. Such strings can be stable, an example of which is an infinite Abrikosov flux tube.[18] In perturbative string theory, type II strings are global due to their coupling to the antisymmetric field $B_{\mu\nu}$.[19]

The fundamental objects in string theory are not point-like, but rather 1-dimensional. This is unlike quantum field theory where local interactions correspond to products of field operators at a point. This includes the creation and annihilation of particles. Since strings, and in particular charged strings, are not point objects, how are they created?

To resolve this creation enigma, it will be assumed here that one end of a charged string is created first and the other a very short time later. It is also assumed that because the string charge has a vector character that the first end always has the same sign dark charge. It is further assumed that string creation occurs, in accord with grand unified theories (GUT), at the time of the symmetry breaking of $SU(5) = SU(3)_C \otimes SU(2)_L \otimes U(1)$ at $\sim 10^{-36}$ sec (the beginning of the standard model symmetry breaking period); i.e., during the period of the GUT phase transitions. It is during one of these phase transitions that inflation, the exponential expansion of the universe, occurs.

As inflation continues, the length of the strings increases exponentially so that a concentration of dark string charge of one sign is rapidly separated from a second concentration of dark string charge of the opposite charge. The first concentration is taken in Fig. 3.1 to reside near infinity and the second concentration to be the dark charges in $\mathbb{S}^3$.

### 5. Evolution of the universe and topological change

The problem with considering $\mathbb{S}^3$ as a model for the very early universe, is that it is now known that the universe is not closed and is either flat or hyperbolic should the matter density be below the critical value even by a small amount. In 1967, Geroch[20] showed that changes in the topology of spacelike sections can occur if and only if the model is acausal. That would seem to rule out $\mathbb{S}^3$. However, Martin, et al.[21] argue that for the FLRW universe quantum topological transitions between curved 3-dimensional hypersurfaces are possible, but are ruled out for a flat hypersurface. This means that classically forbidden topological changes are possible in the quantum domain (see also the earlier work by De Lorenci, et al.[22]). A discussion of these results requires that the Wheeler-De Witt equation and the concepts of superspace, midisuperspace, and minisuperspace be introduced.

The solution to the Wheeler-De Witt equation is a functional—a real valued function on a space of functions; i.e. a function of functions—of positive definite metrics on a 3-manifold evaluated by a Feynman sum over kinematically possible histories. Geroch has given it in the form

$$\Psi(\,^3\mathcal{G}) = \sum_{4-geometries} exp\left[-\frac{i}{\hbar}\int R(-g)^{\frac{1}{2}}d^4x\right].$$

(5.1)

The Wheeler-De Witt equation is obtained by canonical quantization, which could be inconsistent for constrained dynamical systems such as the Einstein gravitational field equations. In addition, when the constraint $H\Psi = 0$ is imposed on the state vector $\Psi$, it is no longer a function of time. Peres[23] has shown that the introduction of a dynamical time can result in a consistent canonical quantization. De Lorenci, et al. introduce a dust field as a time variable to allow a time evolution of the quantum states and this is also used by Martin, et al.

Superspace is the space of geometries for 3-manifolds that constitute *space* in the dynamical picture of general relativity known as geometrodynamics. It can be thought of as the configuration space for general relativity. The associated cotangent bundle can be defined so that it is the phase space for the Hamiltonian formulation. Imposing symmetry restrictions on spacetime metrics leads to minisuperspace where the metrics depend on a finite number of parameters. This turns out to be too restrictive so that one turns to midisuperspace, which results from imposing symmetry requirements on superspace such that the allowed metrics are parameterized by functions rather than numerical parameters.[24]

Martin, et al. assume that the wave functionals obtained from the Wheeler-De Witt equation are of the form $\Psi = e^{iS/\hbar}$, where $S$ is the action, obtained by using the WKB method to solve the equations.

They begin with an FLRW-like metric given by

$$ds^2 = -N^2(t)dt^2 + a^2(t)\left\{d\chi^2 + \left[\frac{\sin(k(t)^{1/2}\chi)}{(k(t)^{1/2}}\right]^2 d\Omega^2\right\},$$

(5.2)

where $d\Omega^2 = d\theta^2 + \sin^2\theta \, d\phi^2$. In this model, topological change occurs when $k(t)$ passes through zero. This metric form can be arrived at by considering the usual FLRW metric,

$$ds^2 = -dt^2 + a^2(t)\left\{\frac{dr^2}{1-kr^2} + r^2 d\Omega^2\right\}.$$

(5.3)

Consider now only the spatial part $dl^2$ and let $\chi = \int \frac{dr}{(1-kr^2)^{1/2}}$, then

$$dl^2 = a^2(t)\{d\chi^2 + f^2(\chi)d\Omega^2\},$$

where

$$f(\chi) = \begin{cases} \sin\chi & (k=1) \\ \chi & (k=0) \\ \sinh\chi & (k=-1). \end{cases}$$

(5.4)

Choosing $k = 1$ and defining $\chi = k^{1/2}\bar{\chi}$ puts the spatial part of the metric into the form

$$dl^2 = ka^2(t)\left\{d\bar{\chi}^2 + \left(\frac{\sin(k^{1/2}\bar{\chi})}{k^{1/2}}\right)^2 d\Omega^2\right\},$$

(5.5)

and making the redefinitions $ka^2(t) \to a(t)$, $\bar{\chi} \to \chi$, and $k \to k(t)$ results in the spatial part of the metric given in Eq. (5.2).

One cannot simply allow $k \to k(t)$ in the FLRW metric, since the metric will then no longer represents a spatially homogeneous spacetime. It is for this reason that De Lorenci, et al., in the paper preceding that by Martin, et al., were forced to introduce a midisuperspace and a metric having a non-vanishing shift function $N(t)$ as shown in Eq. (5.2). This metric is used to determine a Green's function that depends on the volume, $y$, proportional to $a(t)^3$ of the space with 3-curvature $R$, and a dust field $\chi(t)$, which plays the role of time. The semiclassical solution to the Wheeler-De Witt equation is then given by $\Psi = \Psi(y, R, \chi)$.

The Green function $G = (y_f, R_f, \chi_f; y_i, R_i, \chi_i)$ represents the probability amplitude for a topological transition when $R_f$ does not have the same sign as $R_i$ (*i* and *f* designate initial and final values). Computation of this is not possible, so a semiclassical wave function $\Psi = e^{iS/\hbar}$ is assumed in order to find an approximation. For topological change to occur, the phase $S$ must have an imaginary part, allowing topological change to be interpreted as a quantum tunneling effect. As discussed earlier, topological change cannot occur classically. Writing down the explicit Green function found by Martin, et al. would add little to this introductory discussion.

The implication of this is that complex metrics are needed for quantum mechanical changes in topology. Martin, et al. use the metric

$$ds^2 = -N_c^2(t)dt^2 + a^2(t)\{dz^2 + \sin^2 z \, d\Omega^2\}.$$

(5.6)

Giving $N_c$ and $z$ real values results in a positively curved spatial hypersurface; imaginary values give a negatively curved hypersurface. Complexification of the metric changes a positively curved hypersurface to a negatively curved one without introducing a time dependent $k$. In addition, topological change involving flat hypersurfaces are forbidden.

Martin, et al. also found that topological change is improbable for short time intervals and increases as the time interval becomes longer. With increasing time intervals, negative to positive topological changes are suppressed and positive to negative ones are enhanced. They also found that topological changes between large volume spacelike hypersurfaces are very improbable.

The universe is generally assumed to be flat, consistent with the ΛCDM model. Some recent data, however (See Fig. 5.1), indicate that the universe may have a negative curvature.[25] This is due to a hemispheric asymmetry in the cosmic microwave background radiation. See also Sawicki.[26] This possibility, combined with topological change *a là* Martin, et al., opens up the possibility of starting with a closed universe like $\mathbb{S}^3$, which then evolves to a negatively curved universe.

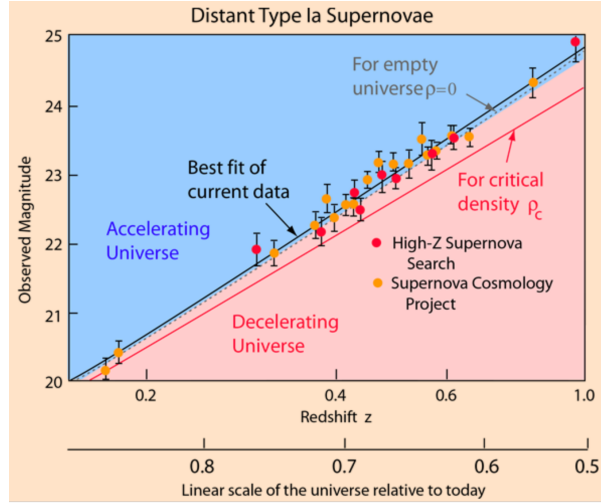

Figure 5.1. Observed magnitude versus redshift plotted for well-measures distant Type Ia supernovae. [Adapted from S. Perlmutter, "Supernovae, Dark Energy, and the Accelerating Universe", *Physics Today*, April 2003.]

## 6. A String Model for Dark Matter[†]

The density profiles of dark matter halos are usually modeled as an approximate solution of the Lane-Emden equation. Using the model proposed here for dark matter coupled with previous work discussed in this section—showing that the approximate solution to the Lane-Emden equation can be an exact solution of the Einstein-Maxwell equations—provides a new insight into the possible nature of dark matter. As above, dark charges and the fields associated with them are assumed to obey Maxwell's equations.

The standard model of ΛCDM has as its principal matter component cold dark matter, which only interacts gravitationally, of an unknown nature. As discussed in the Introduction, relatively recent work on colliding galaxy clusters confirm this supposition[27, 28].

In the case of the rotation curves of galaxies, the density distribution of dark matter is generally assumed to be spherical and to have an isothermal equation of state; i.e., a polytropic equation of state ($P = K\rho^\gamma$) where $\gamma = 1$. The hydrostatic balance equation may then be integrated to yield

$$\rho = \rho_0\, exp(-\Phi/K),$$

---

[†] Much of this section, in the context of electric charge, originally appeared in: G. E. Marsh, "Isothermal spheres and charged dust", *J. Phys. Astron.* **2**, (2013).

(6.1)

where $\Phi$ is the gravitational potential. $\Phi/K$ must then be a solution of the isothermal Lane-Emden equation. Non-singular solutions can be obtained by imposing appropriate boundary conditions, such as requiring that the solution and its first derivative vanish at the origin. The result is an exponential solution for the density of the form

$$\rho(r) = \frac{\rho_0}{exp\left(\frac{\Phi}{K}\right)}.$$

(6.2)

The isothermal Lane-Emden equation cannot be solved analytically and consequently $\Phi/K$ is expanded in a power series. The requirement that the first derivative vanish at the origin limits the expansion to even powers starting with $(\Phi/K)^2$. Expanding the exponential in the denominator of Eq. (6.2), keeping only the first two terms, and using the coefficient given by Chandrasekhar[29] for the leading $(\Phi/K)^2$ term results in the often used expression for the dark matter density,

$$\rho(r) = \rho_0 \frac{r_0^2}{r_0^2 + r^2}$$

(6.3)

where $r_0 = 6K/4\pi G \rho_0$. It will be shown that the right-hand side of this approximate expression corresponds to an *exact* solution of the coupled Einstein-Maxwell equations for dust composed of charged dark matter. Note that if $r_0$ is to be identified with the King radius, the numerical factor of 6 should be replaced by 9.

**7. Dark charged dust**

The term "charge" in this section is meant to designate dark charge rather than electromagnetic charge. The form of the metric for charged dust was introduced by Majumdar [30] and Papapetrou[31]. It is spherically symmetric and static, and can be motivated by considering the Reissner-Nordström metric

$$ds^2 = \left(1 - \frac{2m}{r} + \frac{q^2}{r^2}\right)dt^2 - \left(1 - \frac{2m}{r} + \frac{q^2}{r^2}\right)^{-1} dr^2 - r^2 d\Omega^2.$$

(7.1)

Assume the extreme form of this metric where $|q| = m$, and introduce the isotropic coordinates $\bar{r} = r - m$. Doing so results in the metric

$$ds^2 = f^2\, dt^2 - f^2[dr^2 + r^2 d\Omega^2],$$

(7.2)

where $f = (1 + m/r)^{-1}$. Henceforth the bar above the $r$ will be dropped with the understanding that isotropic coordinates are used in what follows.

Using Newtonian mechanics and classical electrostatics, it is straightforward to show that a system of charged particles of mass $m_i$ and charge $q_i$, where all of the particles have the same sign charge, will be in static equilibrium if $|q_i| = G^{1/2} m_i$. For a continuous distribution of mass $\rho$ and charge $\sigma$, there will be equilibrium everywhere if $|\sigma| = G^{1/2} \rho$. This is what is known as charged dust. It has a general relativistic analog that was discovered by Papapetrou and Majumdar. Although spatial symmetry is not required, spherical symmetry will be assumed here.

Note, however, that the extremal condition $q = G^{1/2} m$ means that if the dark charge $q$ is chosen to be the minimal charge equivalent to one electron or $10^{-19}$ coulomb, then there is a minimal mass of ~$3.6 \times 10^{-9}$ kilogram giving a charge to mass ratio of $2.7 \times 10^{-11}$. This minimal mass is unusual in that it is very close in value to the reduced Plank mass of $\sqrt{\frac{\hbar c}{8\pi G}} = 4.3 \times 10^{-9}$ kilogram (much greater than the supersymmetric extension of the standard model predicting WIMPs having a mass of ~100 Gev/c$^2$).

The equilibrium of charged dust in general relativity has been treated extensively by W.B. Bonnor and others since the early 1960s. It is his paper on the equilibrium of a charged sphere[32] that forms the embarkation point for the work here[33]. The Einstein and Maxwell field equations applied to the metric of Eq. (7.2) show that the Newtonian condition for equilibrium given above must also hold in general relativity. In what follows, the charge will be chosen to be positive.

Bonnor obtained the equation that relates the general form of $f$ to the density,

$$ff'' - 2f'^2 + \frac{2}{r}ff' - 4\pi\rho = 0.$$

(7.3)

Unfortunately, this equation is completely intractable unless $\rho = 0$, and as put by Lemos and Zanchin, "It is not a method for solving the differential equation of the Majundar-Papapetrou problem, it is an art of correct guessing."[34] In other words, one is reduced to guessing a form for the function $f$ and hoping that the equation yields a physically meaningful density distribution.

The problem addressed by Bonnor was to find the density distribution of charged dust within a finite sphere of radius $a$ that would match to the vacuum Reissner-Nordström solution at the boundary. This was successfully achieved using the following expression for $f$

$$f(r) = (a^3 + mr^2)^{1/2}(a+m)^{-3/2}.$$

(7.4)

In Eq. (7.4), $m$ is the mass of the charged dust contained within $r = a$. The density was found to be

$$\rho = \frac{3m}{4\pi a^3}(1+m/a)^{-3}\left(1+\frac{mr^2}{a^3}\right)^{-1}.$$

(7.5)

The question addressed here is whether it is possible to find a function $f(r)$ that would result in a radially unlimited density distribution matching that given in Eq. (6.3) for dark matter. Indeed, one can. Substitution of

$$f(r) = \sqrt{\tfrac{4}{3}\pi\rho_0}\,(a^2+r^2)^{1/2}$$

(7.6)

into Eq (7.3) yields

$$\rho(r) = \rho_0 \frac{a^2}{a^2+r^2},$$

$$\tag{7.7}$$

where $a$ is now a free constant. This has the same form as Eq. (6.3) except that now the equality is exact and $\rho(r)$ is derived from a solution of the Einstein-Maxwell field equations. This is somewhat surprising given that the origins of Eq. (6.3) and Eqs. (7.6) and (7.7) are so different.

Both the isothermal sphere and the corresponding solution given here to the Einstein-Maxwell field equations are unrealistic since the total mass, proportional to $r^{-2}$ at large radii, is infinite. If necessary, it is quite possible that other models can be obtained by modifying these solutions, but galaxies do not really exist in total isolation but rather in galactic clusters so this solution may be adequate.

**Summary**

Some concepts from string theory and quantum topological change in the early universe were used to formulate a new model for dark matter. The end points of open, charged Kalb-Ramond strings are associated with charges, which have been designated here as dark charges where they and their fields obey the Maxwell field equations. This is in keeping with Kalb and Ramond who did not assume that the charges and fields associated with string endpoints are actually electromagnetic in nature, but only that there is an electromagnetic-type interaction between these point charges.

An argument was then given for the presence of dark charges of only one sign appearing at the end of Kalb-Ramond strings in $\mathbb{S}^3$. It was based on the idea of cosmic strings and inflation in the early history of the universe. As the Universe cooled since the Planck time there were a series of spontaneous symmetry breaking phase transitions during which topological defects, such as cosmic strings, could have formed.

Since the fundamental objects in string theory—unlike quantum field theory where local interactions correspond to products of field operators at a point—are not point-like, but rather 1-dimensional, one must come up with a scenario of how they might be created. To do this, it was assumed that in the early history of the universe one end of a charged string is created first and the other a very short time later. It was also assumed that because string charge has a vector character the first end created always has the same sign dark charge. It was then further assumed that string creation occurs, in accordance with grand unified theories (GUT), at the time of the symmetry breaking of $SU(5) = SU(3)_C \otimes SU(2)_L \otimes U(1)$ at $\sim 10^{-36}$ sec (the beginning of the standard model symmetry breaking period); i.e., during the period of the GUT phase transitions. It is during one of these phase transitions that inflation, the exponential expansion of the universe, begins.

Inflation led to the length of the strings increasing exponentially so that a separation of dark string charges occurred. As a consequence, only dark charges of one sign appear in the interior of $\mathbb{S}^3$; the other charges of opposite sign being placed at infinity. It was shown that paired isolating boundary components allow the existence of a single valued Kalb-Ramond potential.

The possibility was then discussed that quantum tunneling could allow $\mathbb{S}^3$ to transition to a negatively curved manifold consistent with what is known of the universe's topology today.

Using the model proposed here for dark matter coupled with previous work discussed in the last section provides a new insight into the possible nature of dark matter.

**Acknowledgement**

I am very grateful to Adam Marsh for his careful reviews of several versions of this paper and for his many helpful suggestions and corrections. The author is of course solely responsible for any remaining errors.

# APPENDICES

## A. Torsion in the bulk and in $\mathbb{S}^3$

As mentioned in Section 1, the field strength, $H_{\mu\nu\rho} = \partial_\mu B_{\nu\rho} + \partial_\nu B_{\rho\mu} + \partial_\rho B_{\mu\nu}$ from string theory is a totally antisymmetric tensor corresponding to a torsion field. This torsion field strength is associated with a string within the bulk, which terminates on a brane, here $\mathbb{S}^3$. If $\mathbb{S}^3$ is to be a hypersurface in a FLRW pseudo-Riemannian manifold there are compatibility constraints on the form of the torsion in $\mathbb{S}^3$. It is interesting that there is experimental evidence from the preferred handedness of spiral galaxies[35,36] that the space we live in today may have a torsion field.

The FLRW constrained form of the torsion in the bulk can be written as

$$T_{\alpha\sigma\nu} = \frac{1}{6} W^\mu \varepsilon_{\mu\alpha\sigma\nu},$$

(A.1)

where $W^\mu$ is a pseudo-vector known as the torsion axial-vector. This means that completely antisymmetric three index tensors are equivalent to pseudo-vectors. If $\alpha = 0$, and the other indices are restricted to spatial coordinates, Eq. (A.1) can be written as

$$T^{0kl} = \frac{1}{6} \varepsilon^{klm} W_m.$$

(A.2)

There is an electromagnetic analogy for the Kalb-Raymon charge and field in string theory. There, the field strength $B_H$ dual to the torsion field $H_{\mu\nu\rho}$ is defined by $\varepsilon^{klm}(B_H)_m = H^{0kl}$, where Latin indices take the values 1,2,3 and $\varepsilon^{klm}$ is totally antisymmetric and satisfies $\varepsilon^{123} = 1$. Note that $H^{ijk} = 0$, $H^{0jk} \neq 0$, and $H$ is time independent (See the book by Zwiebach for a discussion of this ansatz and the review paper by Hammond[37]). The *analogy* with Maxwell's equations is

$$\frac{\partial H^{0kl}}{\partial x^l} = j^{0k} \quad and \quad \vec{\nabla} \times \vec{B}_H = \vec{j}^0.$$

(A.3)

$j^{0k}$ are the charge densities of the string that correspond to the components of a spatial vector that is tangent to the string. This means that the Kalb-Ramond charge density is a vector $\vec{j}^0$ with components $j^{0k}$. The divergence of $\vec{j}^0$ vanishes. A comparison of Eqs. (A.2) and the definition of $\vec{B}_H$ discussed in the paragraph just above Eq. (A.3) shows that Zwiebach's field strength $\vec{B}_H$ is equivalent to $\vec{W}$ and $T^{0kl}$ to $H^{0kl}$.

**B. Single-valued Potentials: Homology, Cohomology, and Cuts**

This Appendix uses the manifold $X = \mathbb{S}^2$ as an example in order to help clarify some of the ideas involved in Section 3. Figure A1 shows a charged Kalb-Ramond string terminating on a 2-sphere.

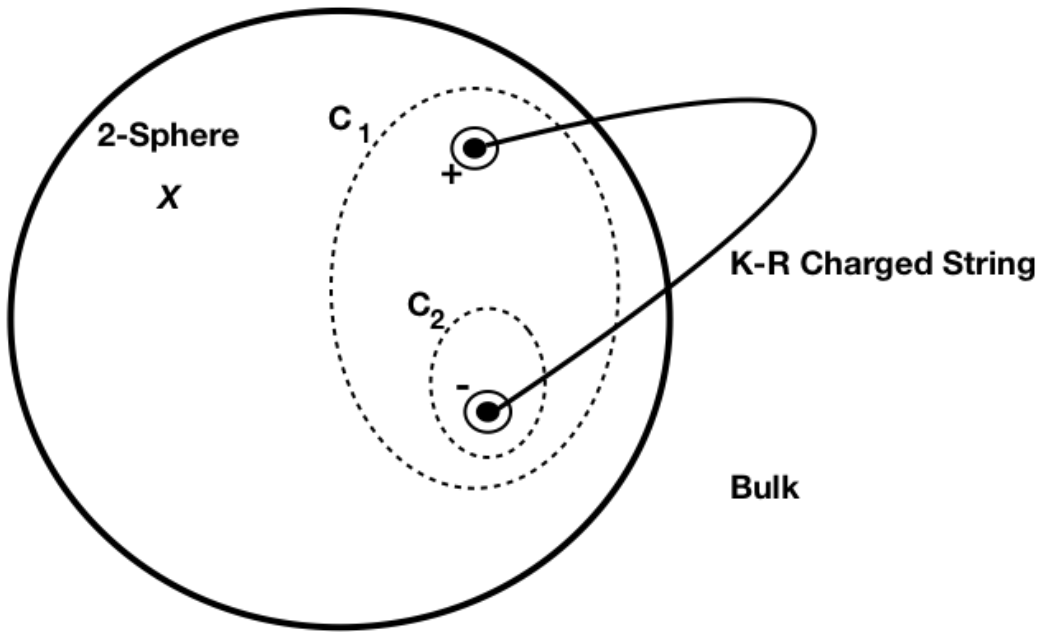

Figure A1. A charged Kalb-Ramond string terminating on a 2-sphere with its dark end charges isolated by boundary components. The closed curve C1 is homologous to zero (because it lies on a 2-sphere), but C2 is not.

In Fig. A1, the bulk is 3-space. Without the string, the homology is given by: $H_0(X) \simeq \mathbb{Z}$; $H_1(X) = 0$; $H_2(X) \simeq \mathbb{Z}$; and $H_3(X) = 0$. With the string terminating on the 2-sphere, this is no longer true since $H_1(X) \neq 0$ because of the obstructions of the boundaries introduced by the charges. A very readable introduction to homology groups can be found in the book by Fraleigh,[38] and a more technical introduction to de Rham cohomology in the book by Warner.[39]

Figure A2 shows how the contractibility of $C_2$ changes with the introduction of a cut (tubular neighborhood) connecting the boundary elements isolating the charges. With the introduction of the cut, $H_1(X) = 0$.

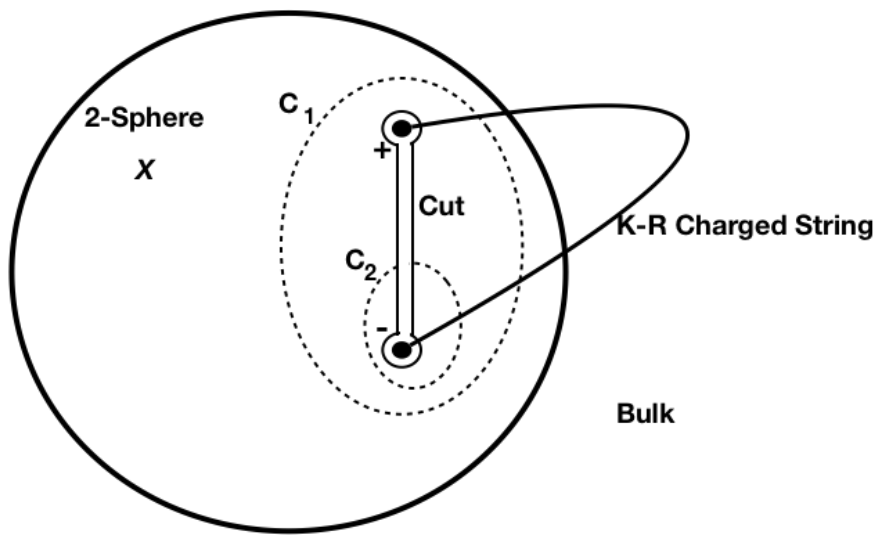

Figure A2. Same as Fig. A1 with the introduction of a cut connecting the two boundary elements isolating the dark charges. The interior of the tubular neighborhood of the cut is in the bulk and the string could just as well now be drawn so as to be contained in this neighborhood. The cut can be shortened vertically so that the positive dark charge and its isolating boundary are contained within $C_2$ so that $C_2$ is also homologous to zero since, as was already the case for $C_1$, $C_2$ is now contractible by going around the 2-sphere.

**Some Background on Forms, de Rham's Theorems, Homology and Cohomology.**

A closed form is one where $d\omega = 0$; an exact form is one for which $\omega = d\eta$; an exact form is also a closed form since $d\omega = d(d\eta) = 0$. Here, $d$ is the exterior derivative. de Rham's first theorem states that a 2-form $d\omega = 0$ is exact if and only if all the periods of $d\omega$ vanish. What this means is that if $\mathbb{M}$ is a manifold and $\sum_i a_i z_i = \partial \mathbb{M}$, where the $z_i$ are boundary components, then $\sum_i a_i \int_{z_i} d\omega = 0$. The integral $\int_{z_i} d\omega$ for each 2-cycle $z_i$ on $\mathbb{M}$ is called a period.

A factor group $G/H = \{Hx | x \in G\}$ is a partition of $G$, as is $G \backslash H = \{xH | x \in G\}$. $Hx$ is the right quotient set and $xH$ the left. $H$ is normal in $G$ if $Hx = xH$ for all $x \in G$; then $G/H = G \backslash H$ and $G/H$ is a group. Let $\phi$ be a mapping from a group $G$ to a group $Q$. The kernel of the mapping is the subgroup of elements in $G$ mapped on to the identity of $Q$. That is, $Ker\phi = \{g \in G | \phi g = 1\}$; the image of $G$ in $Q$ is $Im\phi = \{\phi g | g \in G\}$.

The de Rham cohomology group $H^p_{deR} = \{real\ vector\ space\ of\ closed\ p-forms\}/\{subspace\ of\ exact\ forms\}$; that is,

$$H^p_{deR}(\mathbb{M}, d) = Z^p(\mathbb{M}, d)/B^p(\mathbb{M}, d) = p^{th}\ \text{deRham cohomology group}$$

(B1)

Note that $d^2 = 0 \Rightarrow B^p \subset Z^p$. The dimension of $H^p$ is the $p^{th}$ Betti number, which is finite for compact $\mathbb{M}$. If $\mathbb{M}$ is a smooth manifold, $H^1(\mathbb{M}, d)$ measures the number of holes in $\mathbb{M}$ and $H^0(\mathbb{M}, d)$ measures the number of connected components of $\mathbb{M}$. $H^l(\mathbb{M}, d)$ derives from the sequence of maps

$$\ldots \xrightarrow{d} C^\infty(\mathbb{M}, \Lambda^{l-1}(\mathbb{M})) \xrightarrow{d} C^\infty(\mathbb{M}, \Lambda^l(\mathbb{M})) \xrightarrow{d} C^\infty(\mathbb{M}, \Lambda^{l+1}(\mathbb{M})) \xrightarrow{d} \ldots$$

(B2)

In this equation, $C^\infty$ means that all partial differential equations of all orders exist and are continuous. $\Lambda^p(\mathbb{M})$ is the space of $p$-vectors on $\mathbb{M}$, an $n$-dimensional vector space over $\mathbb{R}$. Note that $H^p(\mathbb{R}^n) = 0\ for\ each\ p \geq 1\ and\ n \geq 1$.

The $l^{th}$ homology group of $\mathbb{M}$ with coefficients in the arbitrary abelian group G is given by

$$H_l(\mathbb{M}, G) = Z_l(\mathbb{M}, G)/B_l(\mathbb{M}, G)$$

(B3)

For the homology groups the sequence of maps is

$$\ldots \xleftarrow{\partial} C_{l-1}(\mathbb{M}, \mathbb{R}) \xleftarrow{\partial} C_l(\mathbb{M}, \mathbb{R}) \xleftarrow{\partial} C_{l+1}(\mathbb{M}, \mathbb{R}) \xleftarrow{\partial} \ldots$$

(B4)

Each $C_l$ is an abelian group and $\partial^2 = 0$.

Consider a manifold $X$ containing an oriented 2-simplex (a triangle with oriented interior) with vertices v₁, v₂, v₃. The boundary map $\partial$ acting on this 2-simplex has as its image an oriented boundary (triangle without its interior). This example gives:

$$Z_l(X, G) = \{c \in C_l(X, G)| \partial c = 0\}\ \text{(cycles)}$$

$$B_l(X, G) = \{\partial c | c \in C_{l+1}(X, G)\} \quad \text{(boundaries)}$$
$$H_l(X, G) = Z_l(X, G)/B_l(X, G) \quad (l^{\text{th}} \text{homology group of X with coefficients in G})$$

(B5)

There is a linear map of the de Rham cohomology to the dual space homology given by $H^p(\mathbb{M}) \longrightarrow H_p(\mathbb{M}, \mathbb{R})^\star$ and defined by

$$\{\alpha\}(\{z\}) = \int_z \alpha.$$

(B6)

In Eq. B6, $\alpha$ is a closed p-form representing the de Rham cohomology class $\{\alpha\}$; $\{z\}$ is a p-cycle representing the real differentiable singular homology class $\{z\}$; and the right hand side of the equation gives real numbers determined by integrals of a differential form over differentiable cycles called the periods of the differential form. The de Rham theorem states that $\star$ is an isomorphism.

- Stokes' theorem states that the periods of an exact form are zero.
- The above isomorphism $\star$ is injective so that the converse will hold: if a closed form has all of its periods equal to zero, then it is an exact form.

Physically, this means that a global potential can be defined.